\documentclass[12pt,a4paper]{article}
\usepackage[T2A]{fontenc}
\usepackage[cp1251]{inputenc}
\usepackage{graphicx}

\begin{document}
\def\a{\it}
\def\bc{\begin{center}}
\def\ec{\end{center}}
\def\lb{\linebreak}
\def\d{\displaystyle}
\baselineskip=18pt
\def\v{\vskip 1 cm}
\def\d{\displaystyle}
\centerline {\bf  The Method of a Two-Point Conditional Column Density}
\centerline {\bf for Estimating the Fractal Dimension of the Galaxy Distribution}
\vskip 0.5 cm
\centerline {\bf Yu. V . Baryshev and Yu. L. Bukhmastova~$^*$ }

\centerline {\it  Astronomical Institute, St. Petersburg State University, Russia}

\centerline {$^*$~bukh@astro.spbu.ru}
\v

{\bf Abstract.}
We suggest a new method for estimating the fractal dimension of the spatial distribution of
galaxies, the method of selected cylinders. We show the capabilities of this method by constructing a twopoint
conditional column density for galaxies with known redshifts from the LEDA database. The fractal
dimension of a sample of LEDA and EDR SDSS galaxies has been estimated to be $D=2.1 \pm 0.1$  for
cylinder lengths of 200~Mpc. Amajor advantage of the suggestedmethod is that it allows scales comparable
to the catalog depth to be analyzed for galaxy surveys in the form of conical sectors and small fields in the
sky. \\


{\bf 1.INTRODUCTION }\\

Various methods are used to statistically analyze the spatial distribution of galaxies. An overview of
these methods is given in the monographs by Martinez and Saar (2002) and Gabirelli et al. (2004),
who, in particular, considered stochastic point fractal processes. 

The method of a conditional number density suggested
by Pietronero (1987) and applied to available galaxy redshift surveys by Silos-Labini et al. (1998)
is a standard method for analyzing the fractal structures
of galaxies. When applied to various samples, this method yields fractal dimensions $D$ of the spatial
galaxy distribution ranging from $1.6$ to $2.2.$ To implement this method requires sequentially selecting
the spherical volumes in the initial sample of galaxies and counting the objects in these spherical volumes.
Since real galaxy surveys often have the form of narrow conical sectors, the method of a conditional
number density is limited by the radius of the largest sphere that completely fits into the volume of a given
survey.

In this paper, we suggest a new method for estimating the fractal dimension-- the method of selected
cylinders. Its advantage over the method of a conditional number density is that it does not require
invoking the spherical volume of a sample, but can be applied to narrow spatial layers, because it uses
information about the distribution of objects along the segments connecting the pairs of points in the
structure.
\v
{\bf 2.  STOCHASTIC FRACTAL STRUCTURES}\\

{\it 2.1. The Density of Stochastic Fractal Structures}\\

The concept of fluid (gas) density, which is commonly used in hydrodynamics, contains the assumption
that there is a density that does not depend on the volume element $dV.$ We can then determine the
density  $\varrho(\vec{x})$ at point  $\vec{x}$ and treat it as an ordinary
continuous function of the position in space. In the problem of analyzing the fluctuations, $\varrho(\vec{x})$ can be
a realization of a stochastic process for which the ordinary moments are specified: the mean, the variance,
etc. In particular, this can be a discrete process containing a finite number of points, for example, a
Poisson process with a particle number density $n(\vec{x}).$ For fractal structures, the concept of particle number
density at a point does not exist, because each volume element of the structure contains an hierarchy
of clusters and the number density depends significantly on the volume element $dV$ (Mandelbrot 1982).
To describe the continuous hierarchy of clustering, which is a new characteristic of the process, we must
introduce a new independent variable-- the radius $r$ of the region in which the particles are counted. In
this case, the number of particles in a self--similar structure increases as a power law
\begin{equation}\label{N-r}
  N(r)=B\cdot r^{D}
\end{equation}
Here, $D$ is the fractal dimension, and $B=N_0/r_0^{D}$ is determined by the number $N_0$ of objects within
the zero--level scale $r_0.$ Two functions are used to characterize the density of fractal structures,
$n_{_V}(r)=N(r)/V(r),$  $n_{_S}(r)=(dN/dr)/S(r),$
$V(r)=(4\pi/3)r^3,~~S(r)=4\pi r^2.$
Let us consider a discrete stochastic process whose realizations represent the sets of points at
random positions $\{\vec{x}_i\}, ~i=1,...N,$  so the realized particle number density $n(\vec{x})$ is given by
\begin{equation}\label{n-x}
  n(\vec{x}) = \sum_{i=1}^N\delta (\vec{x}-\vec{x}_i)\,.
\end{equation}
If we the stochastic process is fractal, then an additional fractal variable $r$ that characterizes the degree
of singularity of the fractal structure must be considered to describe it. Let $N_V(\vec{x}_a, r)$ denote the number
of particles in the sphere of radius $r$ centered at point $\vec{x}_a,$ in the structure:
\begin{equation}\label{N-x-r}
  N_V(\vec{x}_a, r)= \int_0^r n(\vec{x})d^3x\,,
\end{equation}
and $N_S(\vec{x}_i, r)$ is the number of particles in the shell
$r, r+\triangle r$ centered at point $\vec{x}_a$ in the structure:
\begin{equation}\label{S-x-r}
  N_S(\vec{x}_a, r)= \int_r^{r+\triangle r} n(\vec{x})d^3x\,.
\end{equation}
When we pass from one realization to another, these quantities undergo fluctuations; a dependence
on the scale $r$ remains after averaging over the set of realizations. For ergodic processes, the averaging
over realizations may be replaced with the averaging over the points of one realization. According to
Pietroniero (1987), the conditional number density of a stochastic fractal process can be defined as
\begin{equation}\label{eta-r}
  \eta(r) =\left\langle \frac{N_S(\vec{x}_a, r)}{4\pi r^2\triangle r}\right\rangle _{\vec{x}_a} =
  \frac{1}{N}\sum_{i=1}^N \frac{1}{4\pi r^2 \triangle r}
  \int_r^{r+\triangle r} n(\vec{x})d^3x=\frac{{D}B}{4\pi}~r^{-(3-{D})}\,,
\end{equation}
while the volume conditional number density is
\begin{equation}\label{eta-V-r}
  \eta_V(r) = \left\langle \frac{N_V(\vec{x}_a, r)}{(4\pi/3)r^3}
  \right\rangle _{\vec{x}_a}=\frac{1}{N}\sum_{i=1}^N \frac{3}{4\pi r^3}
  \int_0^{r} n(\vec{x})d^3x= \frac{3B}{4\pi}~r^{-(3-D)}
  \,,
\end{equation}

Here, $\langle \cdot \rangle_{\vec{x}_a}$  denotes the averaging performed on the
condition that the centers of the spheres are located at the points occupied by the particles of the realization
(hence the name ''conditional''), and the last equalities
in (\ref{eta-r}) and (\ref{eta-V-r}) apply to the ideal fractal structures (\ref{N-r}), 
for which $\eta_V(r)=(3/{D})\eta(r).$ The exponent in the conditional number density
\begin{equation}\label{gamma}
  \gamma = 3 - D
\end{equation}
is called the fractal codimension of the structure.
A fundamentally important property of the conditional number density is that for processes with finite
fractal scales beyond which the particle distribution becomes uniform, statistics (\ref{eta-r}) and (\ref{eta-V-r}) reach constant
values, which corresponds to the equality $D=3$ for homogeneous structures. Thus, the method of
a conditional number density is a powerful tool in searching for the boundary of the transition from fractal
clustering to homogeneity.
 \v
{\it 2.2. The Two-Point Conditional Column Density}\\

The conditional number densities of stochastic fractal processes considered above are one--point densities,
because the center of the sphere in which the particles are counted is placed at one point $\{a\}$ with
coordinates $\{\vec{x}_a\}.$ In some cosmological problems, for example, those related to gravitational lensing (see,
e.g., Baryshev and Bukhmastova 1997), it is necessary to use two--point conditional number densities where
the two particles $\{a,~b\}$ with coordinates $\{\vec{x}_a,~\vec{x}_b\}$ are
fixed. The transition from the one--point to two--point conditional number densities in the analysis of fractal
structures is similar to the transition from the two--point to three--point correlation functions in the analysis
of ordinary stochastic processes. To characterize the particle distribution along the
cylinder whose axis connects two points of the structure $\{a,~b\}\subset\{x_i\,,\,i=1,...,N\},$  let us introduce the
concept of two--point conditional column density $\eta_{ab}(r)$ of a stochastic fractal process. According to
Mandelbrot's cosmological principle, particles $a$ and $b$ are statistically equivalent; therefore, the one--point
conditional particle number density for each of the points is given by formula (\ref{eta-r}), which is proportional
to the probability of particle occurrence at distance $r$ from the fixed point of the structure. In the case of
two independent fixed points of the structure spaced $r_{ab} =|\vec{x}_a - \vec{x}_b|$ apart, event $C$ that consists in the
occurrence of a particle at distance $r_a$ from point $a$ and the independent occurrence of a particle at
distance $r_b$ from point $b$ of the fractal structure is given by the union  $C=A\bigcup B,$ of the events pertaining to
each of the fixed points $a$ and $b.$ The assumption that events $A$ and $B$ are independent is the first simple
step and can be generalized to dependent events. However, applications of the cylinder method to real
fractal structures have shown that the assumption of independence is a satisfactory approximation.
Let us consider the case where a fractal structure is superimposed on the background of an additional
homogeneous Poisson process. The two--point conditional column density can then be represented as the
sum of the one--point conditional number densities on a Poisson background:
\begin{equation}\label{eta-ab-r}
 \eta_{ab}(r) = \eta_a(r) + \eta_b(r_{ab} - r) + c =
 \frac{{D}B^{'}}{4\pi}\cdot~r_{ab}^{{D}- 3}\cdot
 \left[\left(\frac{r}{r_{ab}}\right)^{{D}- 3} +
 \left(1-\frac{r}{r_{ab}}\right)^{{D}- 3}+c_0\right]\,,
\end{equation}

Here, the distance $r$ is measured along the rectilinear segment connecting particles $a$ and $b$ and simultaneously
determines the radius $r$ of the sphere centered at the first point and the radius $r_{ab}-r$ of the
sphere centered at the second point, and the constant $B^{'}$ can be determined from the normalization condition
that includes the contribution from the homogeneous background specified by the constants  $c$ and $r_0$.
The volume elements in this formula are distributed along the segment $ab,$ and, in this sense, $r$  is a onedimensional
Cartesian coordinate. The following statistics can be used as an estimate
of $\eta_{ab}(r):$  
\begin{equation}\label{eta-ab-r-stat}
  \eta_{ab}(r) =\left\langle \frac{N_c(\vec{x}_a, \vec{x}_b,~r,~h,~\triangle r)}
  {\pi h^2\triangle r}\right\rangle _{\{a,b\}} =
  \frac{1}{N_{ab}}\sum_{\{a,b\}}^{N_{ab}} \frac{1}{\pi h^2 \triangle r}
  \int_r^{r+\triangle r}\int_0^h n(\vec{x})2\pi h ~dh~dr\,,
\end{equation}
where $N_c$ is the number of particles in the volume element of a cylinder with the base radius $h$ and height
$\triangle r,$ and with the axis connecting particles $a$ and $b$ that belong to the fractal structure at distance $r$ from
particle $a,$ which also corresponds to the distance
 $r_{ab} -r$ from particle $b.$ The averaging is over all pairs of points with the cylinders of length $l=r_{ab}\pm m,$
where the distance $m$ is determined by the choice of discreteness in ray length in the sample of objects
under consideration.
\v
{\it 2.3. An Algorithm for Estimating the Two--Point Conditional Column Density}\\

Let us now consider a practical algorithm of the method of selected cylinders to estimate the column
density of objects (galaxies) between specific pairs of points in the structure (fixed galaxies). These
points (galaxy centers) are placed at the centers of the cylinder bases. The cylinder radii are chosen to
be much smaller than the mean separation between the elements of the structure estimated for the sample
volume under consideration. We choose objects with known equatorial coordinates
$\alpha,\delta$ and redshifts $z_G$ from the catalog of galaxies being studied.

{\bf Step 1.} Since a substantial amount of dust is concentrated in the plane of the Galaxy, the number
of galaxies observed through the Galactic equator is significantly reduced. To prevent the distortions introduced
by this observational selection, let us divide our sample into two parts. To this end, we calculate
the Galactic coordinate $b$  of each galaxy in the sample and attribute the galaxies of the Northern ($b>10^{\circ}$) 
and Southern ($b<-10^{\circ}$). Galactic hemispheres to the first and second parts.

{\bf Step 2.}We transform the equatorial coordinates of The galaxies into their Cartesian coordinates:
$\alpha,\delta,z_G\rightarrow x,y,z.$

{\bf Step 3.} We chose any two galaxies and denote
the coordinates of their centers by $(x_1,y_1,z_1)$ and $(x_2,y_2,z_2)$ respectively. We place the points
with these coordinates at the centers of the cylinder bases. We choose the cylinder radius $h=1\cdot 10^{-4}\cdot R_H,$  where $R_H=c/H=5\cdot 10^6$~kpc (below,
$H=60$~km~s~$^{-1}$~Mpc~$^{-1}$ is used).

{\bf Step 4.} We examine all of the remaining galaxies in the sample in turn. Every time, we denote the
coordinates of the galaxy center as  $(x_3,y_3,z_3).$  Following Fig.1, we determine the distance from the
point with coordinates  $(x_3,y_3,z_3)$ to the cylinder axis $h_3=\sin A\cdot a_3\cdot R_H,$ where 
$\cos A=\d{\frac{a_3^2+a_1^2-a_2^2}{2a_3a_1}},$

$$\cases{a_1=\sqrt{(x_2-x_1)^2+(y_2-y_1)^2+(z_2-z_1)^2}\cr
a_2=\sqrt{(x_3-x_1)^2+(y_3-y_1)^2+(z_3-z_1)^2}\cr
a_3=\sqrt{(x_3-x_2)^2+(y_3-y_2)^2+(z_3-z_2)^2}}$$

{\bf The Selection Criterion:} If the galaxy $(x_3, y_3, z_3)$ falls into the cylinder $h_3<h,$ then we calculate the
ratio  $r_{13}/r_{12}=a$(since the base of the selected cylinder is much smaller than its height, we assume that
$r_{13}\sim$  is the projection onto the axis).

{\bf Output data of the method:} the total number $N$ of events when galaxies fall into cylinders, the number
$N(a)$ of events when galaxies fall into certain parts of the cylinder, and $N(a)/N$  is the column density of the
objects between points $(x_1,y_1,z_1)$ and $(x_2,y_2,z_2).$
Let us now write formula  (\ref{eta-ab-r}) as
\begin{equation}
\frac{N(a)}{N}=0.5\cdot\Bigr((R\cdot a)^{-\alpha}
 +(R\cdot(1-a))^{-\alpha}+b\Bigl).
 \label{fmy}
\end{equation}
where  $R$ is the factor defined as the ratio of the maximum length of the segments in the sample to
the minimum fractal scale in the sample, $b$ is the constant corresponding to the background Poisson
distribution, and $\alpha$ is the exponent that determines the fractal dimension $D=3-\alpha.$ The sample should
be analyzed to determine the parameters $R,\,\alpha,\,b.$
If the real distribution of galaxies is fractal, then galaxies statistically ''prefer'' to be clustered toward
the ends of the cylinders located in any two fixed galaxies. Thus, the principal idea of fractality is
demonstrated: all galaxies are equivalent elements of the structure, and the variations in the number
density of the objects with distance from the galaxies follows the same pattern.
\v

{\bf 3. APPLICATION OF THE METHOD}\\

{\it 3.1. A Mock Uniform Distribution}\\

According to formula (8), for a uniform distribution
of objects in space ($D= 3$), the conditional two-- point density is constant. To check whether the
algorithm in this simple case is eficient, we simulated a mock catalog of galaxies with a Poisson distribution
of 2500 points in space. Figure 2 shows the distribution of galaxies between any two fixed points in this
case. It can be immediately seen from this figure that for a uniform distribution of objects in space, there is
no clustering toward the ends of the segments.
\v
{\it 3.2. The LEDA Sample of Galaxies}\\

The LEDA database (Paturel 1997) gives the following astrophysical parameters of galaxies: designations,
morphological descriptions, diameters, magnitudes in various color bands, radial velocities, central
velocity dispersion, etc. The LEDA sample of galaxies contains observational data from various catalogs.
Spatial coordinates are presented for 77 483 galaxies with redshifts $z$ ranging from $10^{-4}$ to $0.25.$ According
to Silos--Labini et al. (1998), the completeness of the LEDA database is: $90\%$ for galaxies with $m_B\leq 14.5$
and known redshifts, $50\%$ for galaxies with $m_B\leq 16$ and known redshifts, and $10\%$ for galaxies with
 $m_B \leq 17$ and known redshifts.

Let us now apply the method of selected cylindersto the LEDA sample of galaxies. We limit the
sample by redshift and absolute magnitude. Consider the volume-- complete samples of galaxies with
$M<-17$ separately for the Northern and Southern Galactic hemispheres. Thus, the number of galaxies
with $z < 0.02 (\sim 100$~Mpc) is 7265 and 6467 in the Northern and Southern parts of the sky, respectively.
The galaxies are fixed in pairs. The points with the coordinates of the centers of these galaxies are located
at the cylinder bases. All of the remaining Northern and Southern sky galaxies are checked to see whether
they belong to these cylinders. We sort the cylinders by height (by segment lengths). How are the galaxies
distributed in the fixed cylinders between two fixed galaxies?

A distinctive feature of a fractal is its self--similarity on any scale. To make sure that this assumption is
valid for the LEDA sample of galaxies, let us single out several subsamples from the total sample containing
the set of segments of various lengths. The first subsample contains all of the segments with
lengths up to 50~Mpc, the second subsample contains the segments with lengths from 50 to 60~Mpc, the
third subsample contains the segments with lengths from 90 to 100~Mpc, and the fourth subsample contains
all of the segments with lengths up to 100~Mpc. Based on formula (8), we can establish the analytical
dependence $N(a)/N$ and estimate the fractal dimension of the distribution on various spatial scales. Table~1 gives the results of our computations.
Figures 3 and 4 show a plot of  $N(a)/N$ for the entire set of segments of various lengths (up to
100~Mpc) for the Southern Galactic Sky on linear (Fig.3) and logarithmic (Fig.4) scales. According to
formula (8), this plot allows us to estimate the fractal dimension of the galaxy distribution in space (i.e., to
determine $D=3-\alpha$).  The exponent $\alpha$ specifies the slope of this function. By varying $\alpha,$ we choose an
optimum value in such a way that the slope of the derived function corresponds to the LEDA data with
the minimum rms dispersion of the fit $\sigma_{fit},$ which is defined as the square root of the sum of the squares of
the difierences between the theoretical (solid line) and observed values of $N(a)/N.$
\v
Table 1
\vskip 0.5 cm
\begin{tabular}{|c|c|c|c|c|c|}
\hline
 Catalog& Number of gal. & Cylinder length, &  $D=3-\alpha$ & $R$ & $\sigma_{fit}$   \\
        & with M<-17     &  Mpc                    &               &     &                   \\
\hline
LEDA\_S&   6467&    0--100   &  2.02 &   243.16 &  0.015 \\
LEDA\_S&    &       90--100  &  2.01 &   245.03 &  0.041 \\
LEDA\_S&    &       50--60   &  1.98 &   216.49 &  0.025 \\  
LEDA\_S&    &       0--50    &  2.18 &   439.64 &  0.028 \\   
\hline
LEDA\_N&    7265&    0--100  &  2.12 &   331.44 &  0.013 \\
LEDA\_N&    &       90--100  &  2.08 &   300.10 &  0.041 \\
LEDA\_N&    &       50--60   &  2.07 &   280.45 &  0.019 \\
LEDA\_N&    &        0--50   &  2.22 &   515.69 &  0.027 \\
\hline
LEDA\_sec 1&  1155  & 50--200 &  2.08 &   288.80 &  0.015 \\
LEDA\_sec 2&   385  & 50--200 &  2.02 &   229.22 &  0.070 \\
\hline
SDSS&    1505&       0--250 &  2.00 &   399.89 &  0.009 \\
\hline

\end{tabular}
\v

An important result is that $b$ is equal to zero in all of the cases considered. This implies that the
observed distribution is purely fractal for all cylinder lengths, and no homogeneous Poisson background
is observed. As expected for fractal structures, the objects are clustered toward the ends of the segments,
and the theoretical dependence (8) well describes the observations without the additional assumption
about the presence of a homogeneous background.

For cylinders of various heights, we obtained dependences similar to those shown in Figs.3 and 4.
This implies that on difierent spatial scales, the galaxy distribution between the pairs of fixed galaxies obeys
the same law  (\ref{fmy}) with a fractal dimension close to $2.$
The fitting parameters $R$ and $D$ in formula \ref{fmy} and $\sigma_{fit}$ for various lengths are listed in the table.

To demonstrate the eficiency of the method for narrow conical sectors, we considered the two corresponding
subsamples of LEDA galaxies presented in the table. The first sector contains 1155 galaxies
with $M <-17,\hskip 0.3cm 0.01 < z < 0.04$ and coordinates $\alpha=143^{\circ}\div 207^{\circ},$ $\delta= 23^{\circ}\div 28^{\circ}.$ Figure~5 shows the distribution
of galaxies in the first sector between the pairs of fixed galaxies on a linear scale.
Sector 2 contains 385 galaxies from the LEDA
database with $M <-17,\hskip 0.3cm 0.01 < z < 0.04$ and coordinates $\alpha=0^{\circ}\div 57^{\circ},$ $\delta= -5^{\circ}\div 0^{\circ}.$  The dependences
for sector 2 are identical to those of sector 1, and the parameters $D,\sigma, R$ are listed in the table.
\v
{\it 3.2. The Sample of EDR SDSS Galaxies}\\

To illustrate the application of our method to a sample of galaxies from a survey in the form of a
narrow conical sector, we considered one band from EDR SDSS (the data were taken from the SDSS
Web page). The table gives the results of our computations of the galaxy distribution in cylinders with
lengths up to 250~Mpc in the sector with coordinates
$\alpha=145^{\circ}\div 235^{\circ},$ $\delta= -1.3^{\circ}\div 1.3^{\circ}.$ 
Figures 7 and 8 show plots of the distribution of 1505 galaxies along the cylinders of all sizes up to
250~Mpc. The fractal dimension was estimated from formula (8) to be $D = 2.00.$ In the latter case, there is
apparently evidence for a homogeneous background.
The constant of the homogeneous background is $b =0.0026.$ However, this issue requires a more detailed
analysis.
\v
{\bf 4. CONCLUSIONS}\\

In this paper, we have suggested a new method for analyzing stochastic fractal structures and showed
that it can be implemented in principle. Our analysis of a sample of LEDA galaxies indicated that the
method of segments with two fixed points (themethod of selected cylinders) can be successfully used to estimate
the fractal dimension of the spatial structures of galaxies in surveys in the form of narrow conical
sections. We used a sample of LEDA galaxies as an example to show that the estimated fractal dimension
of the spatial distribution of galaxies agrees with the value obtained by the method of a one--point conditional
number density (Silos--Labini et al. 1998).

The fractal dimension for the samples of LEDA and EDR SDSS galaxies estimated by our method is
$D= 2.1 \pm 0.1.$ The maximum fractal scale is limited by the length of the segments used in our analysis
(200~Mpc). An analysis of magnitude--limited samples indicated an insignificant influence of selection
effects on the suggested method. The capabilities of the new method will be analyzed in detail in a separate
paper. Note only that the main advantage of the method of a conditional column density is that
it allows scales comparable to the catalog depth to be analyzed for galaxy surveys in the form of conical
sectors and small fields in the sky.
\v
Translated by A. Dambis.
\v
{\large{\bf REFERENCES}}\\

1.{\it Yu. V. Baryshev and Yu. L. Ezova(Bukhmastova),} Astron. Zh. 74, 497,
1997 [Astron. Rep. 41, 436 (1997)].

2. {\it A. Gabrielli, F. Sylos Labini, M. Joyce, and L. Pietronero,} Statistical Physics for Cosmic
Structures (Springer, Berlin, 2004, in press).

3. {\it B. B. Mandelbrot}, The Fractal Geometry of Nature
(Freeman, New York, 1982).

4.{\it V. J. Martinez and E. Saar,} Statistics of the Galaxy Distribution (Chapman \& Hall/CRC, New York,
2002).

5. {\it G. Paturel,} Astrophys. J., Suppl. Ser. 124, 109 (1997).

6. {\it L. Pietronero,} Physica A. 144, 257 (1987).

7. {\it F. Sylos-Labini, M.Montuori, and L. Pietronero,} Phys. Rep. 293, 61 (1998).

\end{document}